# ALGORITHM FOR MULTI-HAND FINGER COUNTING: AN EASY APPROACH

Sumit Kumar Dey[1] and Shubham Anand[2]

[1]Department of Electronics and Communication Engineering, Academy Of Technology, West Bengal University Of Technology, West Bengal, India.
[2]Department of Electronics and Instrumentation Engineering, Netaji Subhash Engineering College, West Bengal University of Technology, West Bengal, India.

## ABSTRACT

*In this paper we propose an easy algorithm for real time hand finger counting involving one or more than one hand. Hand finger counting is a simple medium for Human-Computer Interface which can prove to be a convenient input method for driving interactive menus, small applications and interactive games. Here, we also calculate the orientation of the hand which can be used to provide inputs for the direction and/or speed control of a robot, controlling mouse cursor or slide-show presentations. If being used by a single person, with his or her two hands, the person can trigger ten different commands with fingers in addition to the orientation of the hand. We tried to use very simple algorithm using very basic mathematics of 2D coordinate geometry and avoided the conventional use of calculus, contours and convex hull. Anyone seeking for an easy to implement hand finger counting algorithm can refer to it.*

## 1. INTRODUCTION

Technology has always developed in the direction to simplify our day to day life. Today we are developing towards creating technology even more easy to use. Human Computer Interface (HCI) is the term used to refer the technologies that have been developed for interacting with machines. Till date, a number for hand gesture base HCI have been introduced. Yet here, we want to just add a bit of ease by bringing up an easy to use algorithm that uses computer vision to count fingers of our hands. In an intermediate step we also calculate orientation of our hand which itself can be used for interacting with computer.

## 2. METHOD

We can sub-divide the whole procedure into small manageable fragments described as under.

1. Hand Extraction
2. Noise reduction
3. Calculation of Centroid and orientation

Advances in Vision Computing: An International Journal (AVC) Vol.1, No.1, March 2014

4. Scan for lowest valley
5. Create separate images by finding a split line at lowest valley point
6. Recursively find centroid and orientation of each images and again scan for lowest valley
7. Under no valley condition check if it is a finger

## 2.1. Hand Extraction.

Standard image from the camera consists of coloured pixels in usually RGB format, each of which has a value ranging from 0 to 255 for each of red, blue and green values. This data is stored in a matrix form where each element of the matrix represents a pixel or picture element.

Hand extraction is done by background elimination. [1] It is assumed that the first fame contains only static background. There after the subsequent frames are checked for changes with respect to the first background frame. We use Gaussian blur [2] of 5x5 on the background so as to remove bit of noise and detect sharp changes. We construct a binary image (black & white image) out of the hand that comes over the background. Hand is represented in white on the otherwise black background. We use linear blur of 3x3 and threshold the image such that at least 7 pixels of the concerned 9 pixels have to be white to consider the central pixel to be white. This is a strong way to ensure that no noise speckles arise unless the image is badly affected by noise due to significant movement of camera or in background. This also makes the valleys more prominent and thus they are easier to locate.

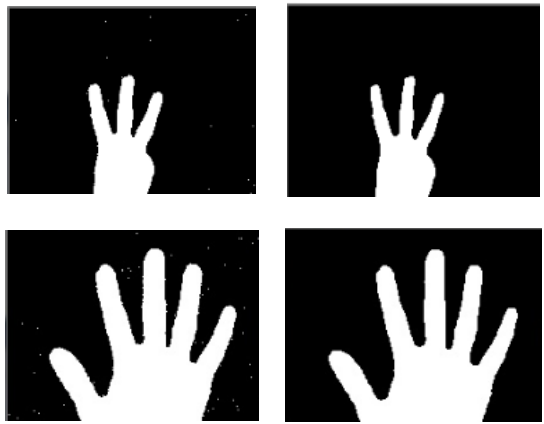

Figure 1. The left binary images are noisy images. The right binary images after application of noise reduction.

The matrix so formed is further reduced by a factor of 3 by spatial sampling in both coordinates to form a much smaller matrix which gives similar image with reduced dimensions. The idea is to replace the bulky matrix with a smaller one so that the calculations can be done in a much faster way involving fewer iterations.

The factor by which the matrix is reduced is a point of trade between speed of the algorithm and the distance of the hand from the camera till where it can detect individual fingers correctly.

## 2.2. Centroid & Orientation

Finding Orientation is an important factor in the algorithm as it helps the algorithm to be valid even if the hand is placed slightly inclined or slanted rather than in the ideal upright orientation. A very simple and basic approach has been followed to find the orientation of the hand.

### 2.2.1. Calculating centroid

We calculate centroid of the white pixels by finding the spatial mean along x and y direction by giving equal unity weights to each pixel.

Centroid(x, y) =   $x_i$ / n,    $y_i$ / n

Where, $x_j$ and $y_i$ represent the x and y coordinates of the pixel in row i and column j. Total number of pixels are represented by n.

### 2.2.2. Calculating orientation

We use a line that gives us the idea about the orientation of the hand. If we show our hand with open palm to the camera such that our middle finger points vertically upwards, then the line perpendicular to our hand, which happens to be the horizontal line in this case, can be termed as base line. Here on, the base line will be used to refer the line which is perpendicular to the hand or to any part of it.

To find the base line, we take the centroid of all points of hand that happen to lie on the lowest horizontal line. Then we find the line joining this centroid to the centroid of the hand that we found in section 2.2.1. Let that line joining the two centroids be represented by $y = m_1.x + c_1$. A line perpendicular to this line will have slope of $(-1/m_1)$. This line, if has to pass through the centroid of lowest horizontal line $(x_{cb}, y_{cb})$, will have the equation as

$$y = (-1/m_1).x + \{y_{cb} + (1/m_1).x_{cb}\}$$

This line intersects the hand almost perpendicularly and here we get our first base line. It roughly replicates the orientation of the hand within the range of ±45º from vertical axis. This orientation can itself serve as a simple way for human computer interaction.

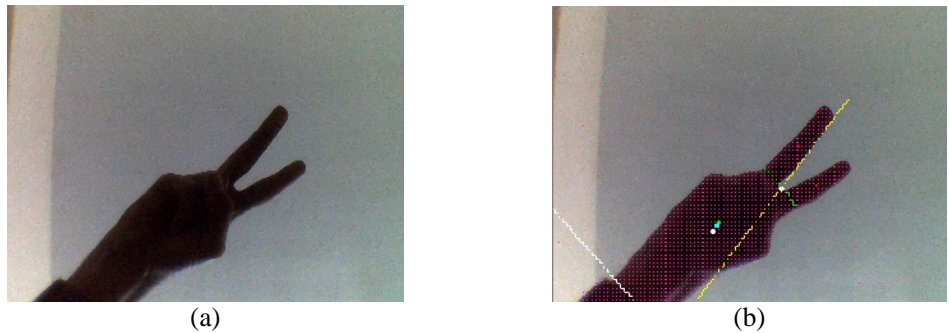

(a)                                                                   (b)

Figure 2. Images depicting orientation. (a) Camera image, (b) the processed image. White line representing the base line.

### 2.3. Scan for lowest valley point

We start scanning the image upwards from the base line mentioned in section 2.2.2 till we find a valley point. Here, we define valley point as spatial mean of the black pixels lying between two white regions. We scan the image along the slope of the base line and for scanning the next upper line we change the y intercept of the line by 1. Thus, we scan the image along a non-conventional axis rather than conventional x-y axis.

The minimum width of the white regions can be fixed according to the expected maximum distance of the hand from the camera. Also, the minimum width of the black region can also be fixed in order to take the minimum spread of fingers in consideration. If minimum width of the black and white regions are not satisfied then it is not considered as a valley point and scanning for valley point is continued. While scanning for a valley point, we turn the scanned pixels to black till we meet the first valley point.

### 2.4. Finding a split line

Upon getting a valid valley point, we must have two disjoint matrices of white pixels. Here, we wish to draw a line such that the two disjoint matrices lie on either of the line. We refer this line as split line, which is a straight line passing through the valley point and has no white pixels on its locus. We find such a line by scanning along the line through valley point starting with the angle 180º and decreasing it gradually down to 0º. We stop scanning as soon as we find a split line. If we do not find any split line upon full scanning then we reject that valley point and again start scanning for new valley point.

On finding a valid split line, we create two copies of the image. Each of the images have white pixels of only one side of the split line. Thus, we separate the two disjoint matrices into two separate matrices. Two points $(x_1, y_1)$ & $(x_2, y_2)$ are on the same side of the straight line $ax + by + c = 0$ if,

$$[ax_1 + by_1 + c \,/\, ax_2 + by_2 + c] > 0$$

Figure 3. Images depicting the centroids, base lines, valley points and split lines

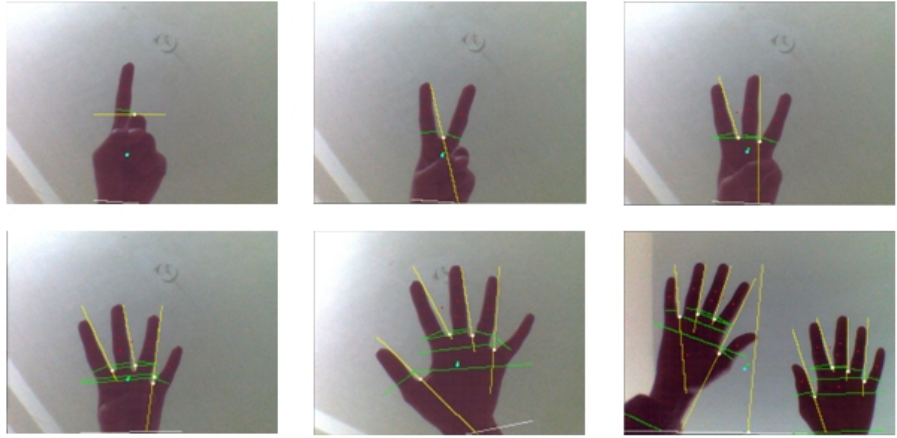

The yellow lines are split lines, Yellow dots are valley points, white lines are the initial or the first base line, the green lines are subsequent base lines and red dots are subsequent centroids.

### 2.5. Going through recursion

Once we get the two separate images, we find the centroids of each of the images. Then we calculate new base line for each of the images. For calculating the new base line we use the base line of the parent image passing through the valley point that produced the separate images and we choose a line slightly higher to it by changing its y-intercept. Along this line we find centroid ($C_n$). The line passing through the valley point and perpendicular to the line joining the centroid of the image and $C_n$ becomes the base line for the new image. Now we scan for valley points in these images. We stop scanning once we cannot find any more valley point.

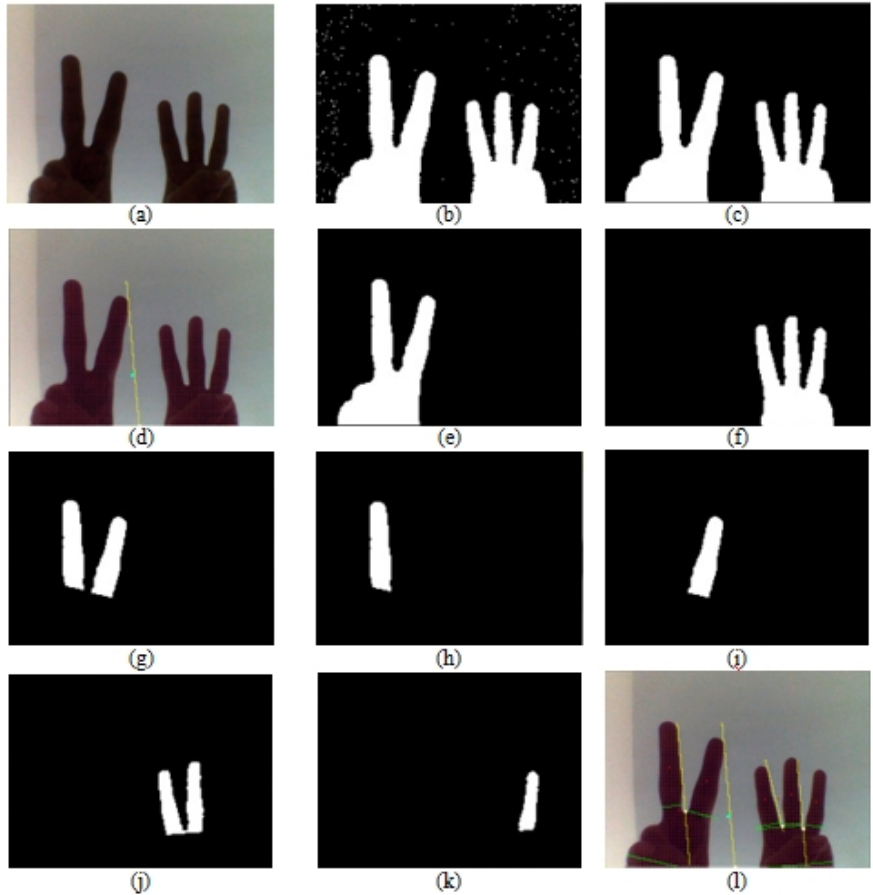

Figure 4. The intermediate steps involved. (a) Image from camera, (b) Binary image with noise, (c) After noise reduction, (d) The split line shown, (e) & (f) are the two separate images from the parent image shown in (c), (g) the binary image at the instant when valley point is found in the image (e), (h) & (i) are the split images of the parent image (e), (j) & (k) are the images from the parent image (f), (l) the image showing all the split lines, base line and valley points after the whole algorithm has executed.

## 2.6. Identifying Fingers

Once we get an image in which no more valley point is present, we need to check if that image represents a single finger or not. At first we reject the images that do not have the certain threshold number of pixels. This threshold number is calculated by squaring the value that was used for minimum number of white pixels in the white region for finding the valley point in section 2.3.

Advances in Vision Computing: An International Journal (AVC) Vol.1, No.1, March 2014

Of the remaining images, we find the number of white pixels lying on the locus of the line parallel to the base line and passing through the centroid. Let the number of pixels found be 'W'. Next we find the number of white pixels lying on the locus of the line perpendicular to the base line and passing through the centroid. Let the number of pixels found be 'H'. We found that we got good results for H/W ratio of more than 1.3.

### 2.7. Performance Analysis

We tested on individuals in different age groups involving different palm sizes. In our tested age range (14 to 52 years) we did not find any accuracy variation on the basis of age but difference in palm size of individuals did affect position of their hands from the camera. Here, under we tabulate the data from 5 users we selected who showed most variations in the readings. User 1 to User 5 are arranged in ascending order of increasing palm size.

Table 1. Accuracy in percentage for different users for all finger counts

| Finger Count | User 1 | User 2 | User 3 | User 4 | User 5 | Overall |
|---|---|---|---|---|---|---|
| 1 | 100.0 | 99.0 | 99.3 | 97.6 | 98.6 | 98.89 |
| 2 | 96.4 | 100.0 | 98.8 | 84.3 | 92.6 | 94.41 |
| 3 | 97.0 | 84.6 | 97.0 | 96.8 | 96.6 | 94.39 |
| 4 | 100.0 | 100.0 | 99.6 | 98.8 | 93.8 | 98.34 |
| 5 | 96.0 | 96.6 | 98.0 | 96.0 | 97.6 | 96.84 |
| 6 | 87.2 | 86.6 | 95.0 | 88.4 | 92.8 | 90.00 |
| 7 | 84.0 | 87.0 | 90.0 | 88.6 | 94.4 | 88.81 |
| 8 | 95.2 | 93.8 | 96.0 | 98.8 | 95.4 | 95.83 |
| 9 | 93.6 | 90.2 | 86.4 | 89.2 | 92.2 | 90.32 |
| 10 | 98.0 | 97.6 | 95.3 | 97.8 | 97.2 | 97.18 |
| Mean for all Users | 94.74 | 93.54 | 95.53 | 93.61 | 95.12 | |

Due to the auto white balance adjust of the camera, we got a lot of noise when posing with both hands. Thus, users had to adjust the distance of their hands from the camera so as to minimise the changes due to white balance adjust. Our code provided visual feedback of number of fingers being detected in each frame, so users could easily adjust their hands to get the best results.

In most cases we got a few erroneous results in first few frames when the hand was slowly entering the view field of camera. This was because of the fact that all the fingers were not at once inside the view frame of camera. So, the frames captured while the hand was moving inside the view field of camera yielded erroneous detection results.

All the errors that we found occurred due to two reasons as mentioned above - auto white balance adjust and frames in which the hand was not completely in the view field of the camera. If we neglect the effect of auto white balance, the distance of hand from the camera did not cause any significant problem to any user. Usually, all users put their hands at a distance in between 30cm to 100cm from the camera.

### 3. CONCLUSIONS

We intended to develop a very easy algorithm and we were successful in our endeavour. The algorithm has great amount of flexibility and thus can be used for numerous applications with accurate results.

Advances in Vision Computing: An International Journal (AVC) Vol.1, No.1, March 2014

We proceeded with few assumptions. Firstly we assume that the hand is shown from bottom of the image such that fingers lie on relatively upper part of the image with respect to the rest of the hand. Secondly, we assume that whenever the hand comes into view, it shows fingers and not a closed fist.

We successfully used the algorithm to count up to ten fingers under the resolution of 640x480. If not constrained by the resolution of the camera we can have any number of hands for finger counting.

The code was written in C using Visual Studio 13 with OpenCV. The algorithm was tested on a machine featuring AMD Quad-Core A10 2.5GHz processor and 8GB RAM on Windows 7 platform. We used Creative VF0520 Live! Cam Sync webcam having USB interface.

It can primarily be used for Human-Computer Interaction. For example, controlling mouse cursor, slide shows and can even be used for controlling a robot. By using our two hands we can trigger 10 unique commands and by changing the orientation of our hands we can provide even more commands like changing the direction of movement of a robot or speed control of robot.

## ACKNOWLEDGEMENTS

We would like to thank Pijush Kar, Sourabhmoy Bhowmick, Soumen Sadhukhan, Jyotirmoy Dendria for their immense support and encouragement.

A special thanks to Mr Sujoy Mondal for his suggestions and valuable advice.

We are profoundly grateful to Mr Manoj Kumar De for his constant guidance and support whenever we required and without whom this paper would never be possible.

We would like to thank our parents and teachers whose very existence mattered significantly for our success.

## REFERENCES

[1]  Ankit Gupta and Kumar Ashis Pati, A Project on Finger Tips Detection and Gesture Recognition, Indian Institute of Technology, Kanpur. November 12, 2009
[2]  Gary Bradski and Adrian Kaehler, September 2008, Learning OpenCV, O'Reilly, Shroff publishers and distributers.
[3]  Shaikh Shabnam Ahmed, Shah Aqueel Ahmed and Sayyad Farook Bashir, (2013) International Journal of Engineering Research & Technology (IJERT), Vol. 2, Issue 10, October 2013
[4]  V. Bansal, S. Singh, H. Arora and S. Tiwari, Project on Simple Cloth Modeling and Control using Finger Count, Jaypee Institute of Information Technology, 2012
[5]  Stephen C. Crampton and Margrit Betke, Counting Fingers in Real Time: A Webcam based Human Computer Interface with Game Applications, Boston University, from http://citeseerx.ist.psu.edu/viewdoc/download?doi=10.1.1.69.9474&rep=rep1&type=pdf
[6]  Daeho Lee and SeungGwan Lee, Vision-Based Finger Action Recognition by Angle Detection and Contour Analysis, ETRI Journal, Vol. 33, Number 3, June 2011
[7]  Mokhtar M. Hasan and Pramod K. Mishra, Novel Algorithm for Multi Hand Detection and Geometric Features Extraction and Recognition, International Journal of Scientific & Engineering Research Vol 3, Issue 5, May 2012
[8]  Hojoon Park, A Method for Controlling Mouse Movement using a Real-Time Camera, Brown University, from http://cs.brown.edu/research/pubs/theses/masters/2010/park.pdf

**Authors**

**Sumit Kumar Dey**
He is currently pursuing Bachelor of Technology in Electronics and Communication Engineering from Academy of Technology (WBUT), Kolkata. A passionate programmer interested in Computer Vision and Robotics.
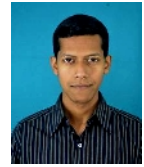

**Shubham Anand**
He is currently pursuing Bachelor of Technology in Applied Electronics and Instrumentation Engineering from Netaji Subhash Engineering College (WBUT), Kolkata. A robotics enthusiastic with deep interested in Computer Vision.
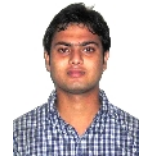